# Current Status of Hard X-Ray Nano-Tomography on the Transmission Microscope at ANATOMIX Beamline

**Mario Scheel[1, a)], Jonathan Perrin[1], Frieder Koch[2], Guillaume Daniel[1], Jean-Luc Giorgetta[1], Gilles Cauchon[1], Andrew King[1], Viktoria Yurgens[2], Vincent Le Roux[1], Christian David[2], and Timm Weitkamp[1]**

[1] Synchrotron SOLEIL, Gif-sur-Yvette, France.
[2] Paul Scherrer Institut, Villigen, Switzerland.

a) mario.scheel@synchrotron-soleil.fr

The transmission X-ray microscope (TXM) on the Anatomix beamline welcomed its first nano-tomography users in 2019. The instrument is based on diffractive optics and works in the range of energies from 7 keV to 21 keV. A spatial resolution in 3D volumes of better than 100 nm can be achieved. The design allows imaging samples in air, and local tomography as well as off-axis tomography scans are possible. Scans below and above K-edges can be made to access elemental distribution. The TXM serves materials science and the bio-medical field.

**1. Design of the TXM**

The hard X-ray full-field microscope is one main instrument of the imaging beamline ANATOMIX at Synchrotron SOLEIL [1, 2]. The transmission X-ray microscope (TXM) is based on diffractive optics and the design energy is in the range from 7 to 11 keV and it allows two fixed energies of 17 keV and 21 keV. The instrument yields a spatial resolution below 100 nm and pixel sizes below 25 nm.

The diffractive optics for the TXM were manufactured at the Laboratory for Micro- and Nanotechnology of the Paul Scherrer Institute in Villigen, Switzerland. The optics have so far mainly been tested at 10 keV where the beam shaper [3] (diameter 2.5 mm, smallest zones of 50 nm) illuminates a field of view of $40 \times 40$ µm². The objective zone plate (diameter 100 µm, smallest zones of 50 nm) has a focal distance of 40 mm at 10 keV. The diffractive X-ray optics were made of iridium using the frequency-doubling method [4], with structure heights of about 1 µm for the beam shaper and 1.6 µm for the objective zone plates, which were patterned on both sides of the membrane [5, 6].

The other elements of the TXM instruments are a rotating diffuser (paper sheet) mounted at the entrance, an order-selecting aperture (pinhole of 0.5 mm diameter) at a distance of 120 mm from the sample position. A phase mask consisting of an array of phase shifting rings can be placed after the objective zone plate to obtain for Zernike-type phase contrast. The main detector, located at 1.3 m or at 2.5 m from the sample, is a CMOS-based digital camera with 6.5-µm pixels (Hamamatsu Orca Flash 4 V2) coupled to indirect microscope optics (tube lens, mirror, 10x Mitutoyo objective, LuAG scintillator). When using a 2×2-pixel binning of the camera the resulting effective pixel size is 50 nm. The detector optics is mounted on standard translation stages (Huber Diffraktionstechnik, Rimsting, Germany), the sample translation stages are designed by SOLEIL and the TXM optics uses piezo-based translation stages (SmarAct, Oldenburg, Germany). The main tomography scan axis is implemented via a high-precision rotation stage (model RT150U, LAB Motion Systems, Leuven, Belgium) whose runout had previously been characterized at SOLEIL to be below 20 nm.

The back side view of the TXM is shown in **FIGURE 1.** The X-ray beam enters (left side of the picture) via a Kapton window (not visible in picture) into the condenser chamber which is filled with $He_2$ gas to reduce the X-ray absorption. In the open space between the chambers (central area in picture) the sample stage is mounted. Downstream of the sample position, another chamber (partly visible at right) houses the objective zone plate and phase mask. This chamber can also be filled with $He_2$ gas. A plastic curtain (visible in the background of the picture) surrounds the instrument, provides thermal insulation from the rest of the hutch and reduces the number of dust particles.

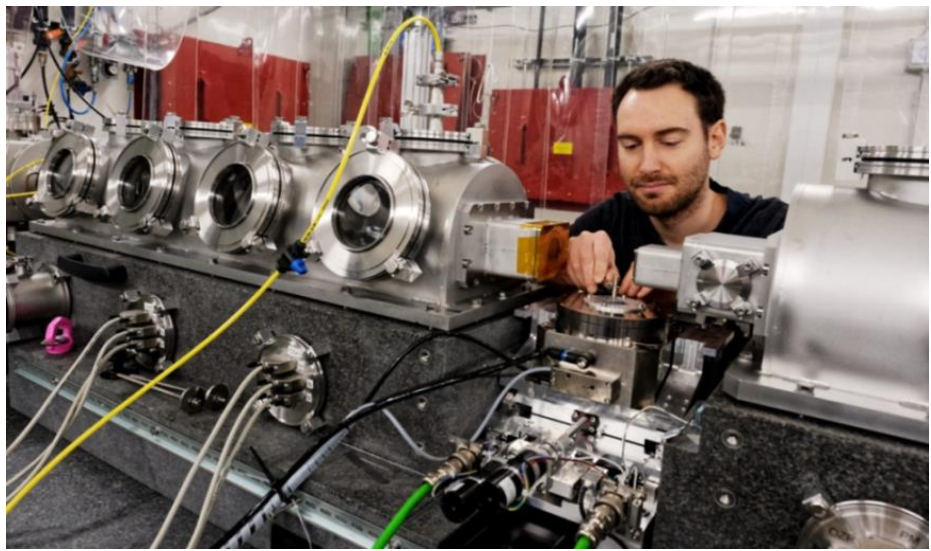

**FIGURE 1.** Backside view of the nano-tomography station of the transmission X-ray microscope. The X-ray beam arrives from the left side and enters into the condenser chamber (with four viewports). In the space at the middle the samples stages are mounted with a rotation stage on the top. On the right, part of the objective chamber can be seen.

## 2. Performance of the TXM

The TXM already performed well during first tests, even though it was not fully completed. For these tests, the cryogenically-cooled, in-vacuum U18 undulator source was used at 7.25 mm gap. The monochromatic X-ray beam of 10 keV from a double crystal monochromator (DCM, Si-111) was collimated with a beryllium compound refractive lens, located at 37 m from source, onto the condenser of the instrument at 168 m from source. **FIGURE 2** (a) shows a TXM nano-radiograph obtained with this setup on a resolution test chart (model XRESO-20, NTT-AT, Japan; tantalum structures of 100 nm height). The smallest structures in the center of this reference sample of "Siemens-star" type in the image (see inset of **FIGURE 2** (a)) have a structure width of 15 nm as measured by NTT-AT, and are well below the Nyquist limit of the resolution for the pixel size of 10.5 nm. The estimated resolution of our measurement of 25 minutes was about 25 nm (half pair line).

Using the same setup **FIGURE 2** (b) shows a Zernike phase contrast nano-tomography of a cement sample which was prepared with FIB-SEM (i.e., a focused-ion-beam milling combined with scanning electron microscopy). The vertical reconstructed slice (i.e., a slice parallel to the tomography rotation axis) of a tomography volume was obtained in 100 minutes at a pixel size of 22 nm (500 projection angles over 180°, 10 s exposure time per projection micrograph, superposition of 6 scans of this type). The resolution in the tomogram is estimated to be around 80 nm, limited mostly by mechanical drift due to temperature variations during the scan.

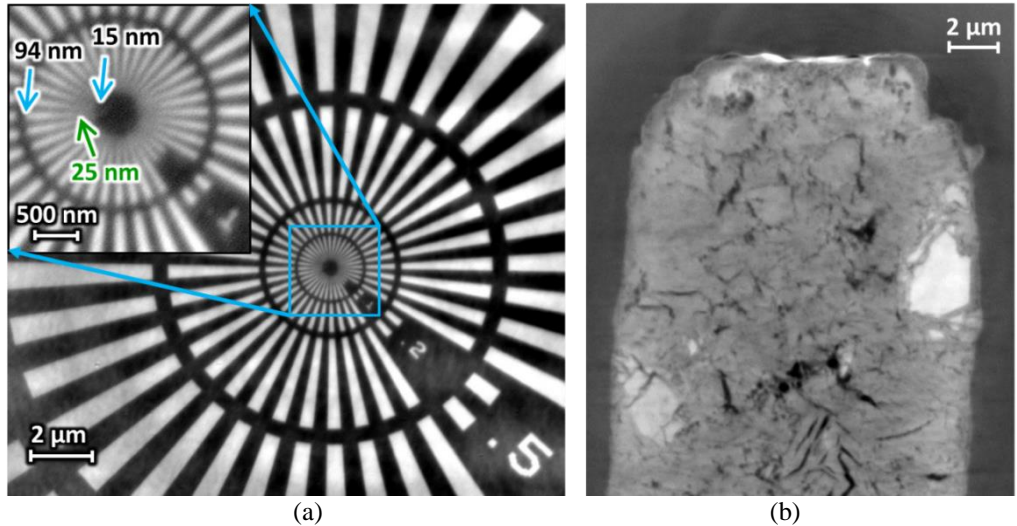

**FIGURE 2.** (a) TXM nano-radiograph of a resolution test pattern. (b) Vertical slice of a nano-tomography of a cement sample.

The TXM instrument shows a good thermal stability which allows for local Zernike nano-tomography as long as the time for a tomography is below 20 minutes and if the sample size induces a phase shift below π. During the first user experiment we combined micro-tomography with nano-tomography of the same seed of *Arabidopsis thaliana* in the same user experiment. The user team studied the Columbia ecotype (col-0, seed diameter around 0.3 mm) **without physically cutting the specimens or applying staining protocols to enhance contrast. FIGURE 3** shows virtual cuts extracted from 3D volume images obtained with tomography at different length scales up to the full grain. In **FIGURE 3** (a) the seed is shown using micro-tomography whereas in **FIGURE 3** (b) the local Zernike nano-tomography at a pixel size of 95 nm of the same seed and height is shown. The reconstructed volume size can be increased with off-axis tomography (rotation axis displaced by 15 μm), resulting here in a reconstructed region of 70 μm diameter and 40 μm height. Several scans of the same region were made; the two horizontal translation stages on top of the rotation stage can be used to acquire different regions of interest in the sample whose volume images can subsequently be stitched together. During the four days of the experiment, the user team acquired more than 200 nano-tomography and high-resolution micro-tomography volume scans.

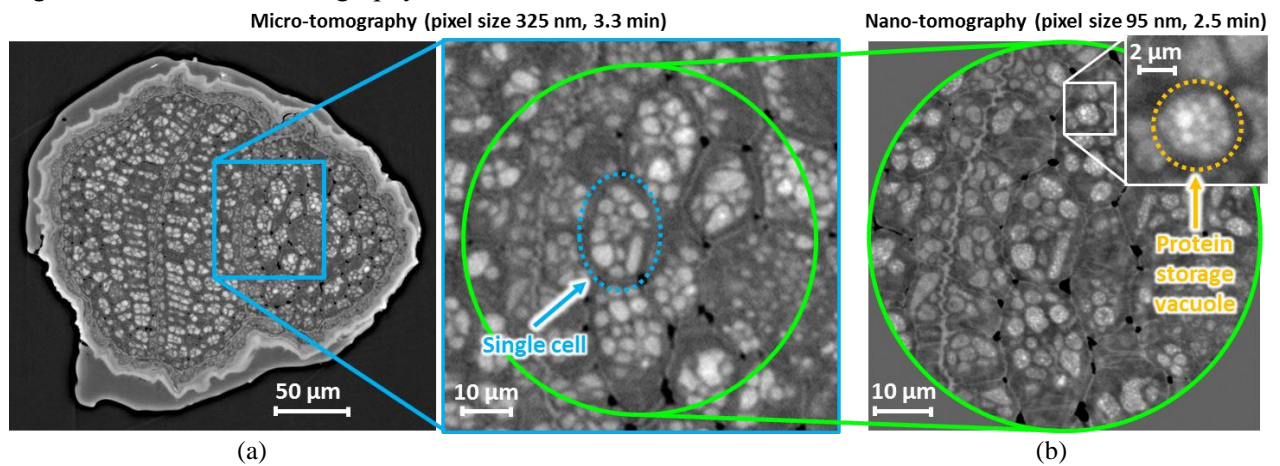

**FIGURE 3.** (a) Tomography images (virtual slices) of Arabidopsis thaliana, Columbia ecotype (col-0), obtained on ANATOMIX in micro-tomography (left: full grain, center: detail) and (b) TXM nano-tomography (right). Note the organelles visible in the nanotomography image (inset on far right).

In some cases the TXM can be used for extracting chemical information, for example if the K-edge of an element is accessible. **FIGURE 4** (a) shows three vertical slices of nano-tomography volumes acquired at three energies around the Cu K-edge. The outer shell of the sample is mainly made of copper and shows a strong change in X-ray absorption depending on the X-ray photon energy. In **FIGURE 4** (b) the result of a subtraction of images taken 7 eV below and 7 eV above the Cu K-edge is shown. The TXM was used in absorption contrast at a pixel size of 40 nm. The estimated resolution is 150 nm. The total exposure time for one energy scan was 67 minutes (four scans were aligned and averaged). The sample had been prepared by focused-ion-beam milling combined with scanning electron microscopy (FIB-SEM).

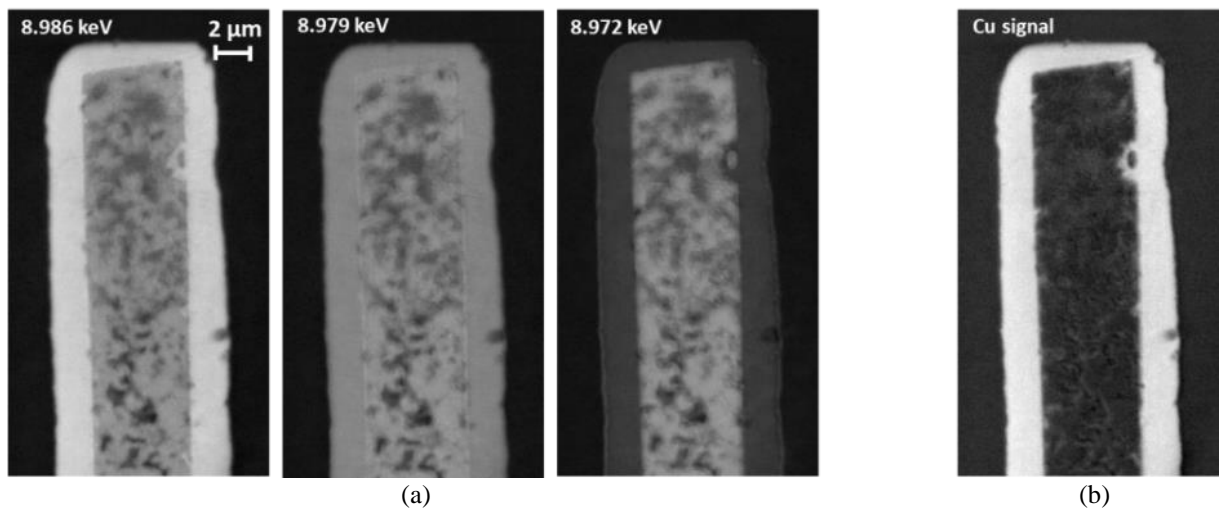

(a)          (b)

**FIGURE 4.** (a) Vertical slices of nano-tomographies of a sintered Ag sample with Cu cladding, taken at three X-ray photon energies close to the Cu K-edge. (b) Vertical slice through the subtraction image 7 eV below and above the Cu K edge.

## 3. Conclusion

The first TXM user experiments confirmed the good performance and stability of the instrument. A 3D resolution below 100 nm can be achieved at scan times of 30 minutes. At larger pixel sizes, in 2×2 binning mode, the spatial resolution is degraded to 150 nm but the scan time is decreased to below 5 minutes. Absorption contrast as well as Zernike phase contrast are provided, and we offer the options to use local tomography or off-axis tomography for enlarging the field of view. This means that relatively large samples can be measured, which in many cases avoids complicated and time consuming sample preparation. The fast scan times of a few minutes enable studies in which many samples have to be scanned, such as biological or medical samples. In addition, micro-tomography scans and nano-tomography scans can be acquired in the same user session. This unique TXM offers several magnifications via short detector options of 1.3 m or 2.5 m as well a long option of 30 m into a second experimental hutch.

## 4. Acknowledgments
ANATOMIX is an Equipment of Excellence (EQUIPEX) funded by the Investments for the Future program of the French National Research Agency (ANR), project NanoimagesX, grant no. ANR-11-EQPX-0031. The authors thank Camille Rivard and Yann Gohon for the Arabidopsis seeds, Roland Pellenq for the cement sample, and Yannick Pannier and Xavier Milhet for the sintered AgCu sample.